\documentclass[prb,twocolumn,psfig,show pacs]{revtex4}
\usepackage{amsfonts}
\usepackage{amsmath}
\usepackage{graphicx}
\usepackage{bm}
\usepackage{amssymb}
\usepackage{ulem}
\usepackage{times}
\usepackage{dcolumn}
\usepackage{cases}
\usepackage{txfonts}


\begin{document}

\title{Optimized Decimation of Tensor Networks with Super-orthogonalization
for Two-Dimensional Quantum Lattice Models}
\author{Shi-Ju Ran, Wei Li, Bin Xi, Zhe Zhang and Gang Su}
\email[Corresponding author. ]{Email: gsu@gucas.ac.cn}
\affiliation{Theoretical Condensed Matter Physics and Computational
Materials Physics Laboratory, School of Physical Sciences, Graduate
University of Chinese Academy of Sciences, P. O. Box 4588, Beijing
100049, China}

\begin{abstract}
A novel algorithm based on the optimized decimation of tensor
networks with super-orthogonalization (ODTNS) that can be applied
to simulate efficiently and accurately not only the thermodynamic
but also the ground state properties of two-dimensional (2D) quantum
lattice models is proposed. By transforming the 2D quantum model
into a three-dimensional (3D) closed tensor network (TN) comprised
of the tensor product density operator and a 3D brick-wall TN, the
free energy of the system can be calculated with the imaginary time
evolution, in which the network Tucker decomposition is suggested
for the first time to obtain the optimal lower-dimensional
approximation on the bond space by transforming the TN into a
super-orthogonal form. The efficiency and accuracy of this algorithm
are testified, which are fairly comparable with the quantum Monte Carlo
calculations. Besides, the present ODTNS scheme can also be
applicable to the 2D frustrated quantum spin models with nice
efficiency.
\end{abstract}

\maketitle
\section{Introduction}

Efficient and accurate numerical methods are very crucial to tackle the strongly correlated quantum lattice systems. To a large class of intriguing correlated electron and spin models, analytical techniques are intractable owing to their extreme complexity and meanwhile, numerical approaches are still challenged by the huge Hilbert space that increases exponentially with the lattice size. Two decades ago, numerical renormalization group algorithms based on density matrix for the ground states \cite{DMRG} and thermodynamic properties \cite{TMRG} of one-dimensional (1D) systems were proposed, where the thoughtful selection rules were suggested for optimally approximating the Hilbert space with an effective subspace.
Very recently, efficient representations with tensor networks as
well as the corresponding algorithms for the two-dimensional (2D) quantum
models, for instance, the projected entangle pair state (PEPS)
\cite{PEPS}, the tree tensor network  \cite{TTN}, the multiscale
entanglement renormalization ansatz state \cite{MERA}, the infinite
PEPS \cite{projection, projection2}, the tensor renormalization
group (TRG) \cite{TRG1, TRG2, TRG3}, and so on, have been suggested. Some
of them already gained interesting applications (e.g. Refs.
[\onlinecite{Li, Chen}]). These algorithms are well testified for calculating the ground
state properties, while the algorithms for the
thermodynamics of the infinite 2D quantum models still need to be developed.

In this paper, we propose the optimized decimation of tensor
networks with super-orthogonalization (ODTNS) to simulate efficiently
not only the thermodynamic but also the ground state properties of 2D
quantum spin lattice models. Inspired by the projection method of the ground states of 2D systems \cite{TRG2} and the linearized TRG method for thermodynamic properties of 1D systems \cite{LTRG}, we represent the finite temperature density operator of the 2D quantum model with a
three-dimensional (3D) closed tensor network (TN) that consists of the initial tensor product
density operator (TPDO) and the 3D brick-wall TN for the evolution
along the imaginary time direction. The finite temperature
properties can be obtained by linearly contracting the brick-wall
TN with the corresponding imaginary time length to get the TPDO
\cite{LTRG}. To bound the dimension of the TPDO, we develop the Tucker decomposition \cite{TD} to the TN and propose the network Tucker decomposition (NTD) that transforms a TN into the
super-orthogonal form so that an optimal lower-dimensional
approximation for the bond space can be reached based on the network singular value spectrum.
We testify the efficiency of the ODTNS scheme by calculating the
thermodynamic properties of the  unfrustrated spin-$1/2$ Heisenberg
antiferromagnet on single-layer and bilayer honeycomb lattices, and the obtained results
show the great agreement with quantum Monte Carlo (QMC)
calculations. We also calculate the thermodynamic and magnetic properties of the frustrated bilayer model to study the effect of frustration. In what follows, we shall present the procedure of the ODTNS algorithm with a 2D quantum spin system on a honeycomb lattice as a prototype.

\section{Tensor network representation of the finite temperature density operator}

\begin{figure}[tbp]
\includegraphics[angle=0,width=1\linewidth]{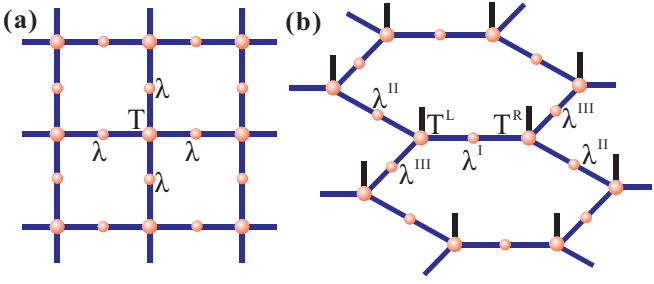}
\caption{(Color online) (a) A closed square TN in which each tensor
$T$ has four geometrical bonds that connect each other and no physical bond.
On each geometrical bond there defines a vector $\lambda$. (b) An open honeycomb
TN consisting of two inequivalent tensors $T^L$ and $T^R$, each of which has
three geometrical bonds and one physical bond. Three inequivalent vectors
$\lambda^I$, $\lambda^{II}$ and $\lambda^{III}$ are defined on three
inequivalent bonds of the TN, respectively.}
\label{fig-TN}
\end{figure}

In accordance with the general definition of the TN's, we define a TN as a network consisting of the
product of tensors ($T$) and vectors ($\lambda$), as shown in Fig.
\ref{fig-TN}. Graphically, a point with some connected bonds
represents a tensor; each bond represents an index; a bond that
connects two points is called geometrical bond which means a shared
index by two tensors that should be contracted; a bond that only
connects one point is called the physical bond. We restrict here
that one bond must connect one or two points. If an index is shared
by more than two tensors (say $n$), the restriction would always be
fulfilled by introducing an $n$th-order super-diagonal tensor. The
points on the geometrical bonds represent vectors. A TN with no
physical bonds is called a closed TN [Fig. \ref{fig-TN} (a)], e.g. a
TN that denotes the partition function of a classical model; a TN
with N physical bonds is called an open TN, which contains $d^N$
degrees of freedom (where $d$ is the dimension of one physical bond,
Fig. \ref{fig-TN} (b)), e.g. a TN that represents a tensor product
state or a tensor product operator.

\begin{figure}[tbp]
\includegraphics[angle=0,width=1\linewidth]{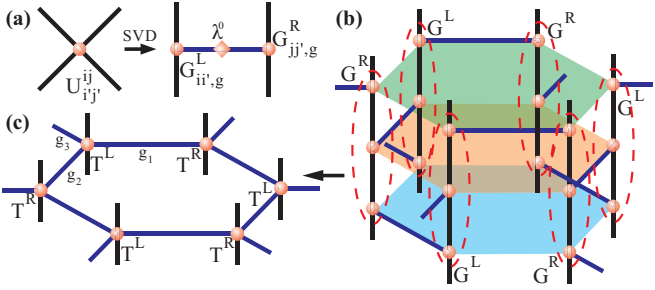}
\caption{(Color online) (a) The local evolution operator
$U^{ij}_{i'j'}$ is decomposed via an SVD into two gates,
$G^{L}_{ij,g}$ and $G^{R}_{i'j',g}$, each of which has two physical
bonds ($i,i'$ and $j,j'$, black) and one geometrical bond ($g$,
blue); (b) Contract the shared physical bonds among $G^L$ and $G^R$
to get tensors $T^L$ and $T^R$; (c) A TPDO with inverse temperature
$\tau$. Note that the singular value vectors $\lambda^{I,II,III}$ on
each geometrical bond are not indicated in (b) and (c) for conciseness.}
\label{fig-TensorNet}
\end{figure}

The finite temperature density operator of a 2D system can be transformed into an open TN.
Suppose that the Hamiltonian can be written as
$H=\sum_{i,j}\hat{H}_{ij}$, where $\hat{H}_{ij}$ is a local
Hamiltonian of pairs of spins. The partition function $Z$ is the
trace of the density matrix $\rho = \exp (-\beta H)$ with
$\beta=1/T$ the inverse temperature and $k_B=1$. By means of the
Trotter-Suzuki decomposition \cite{Trotter}, the density operator can be written as
$\rho \simeq [\exp{(-\tau \sum_{i,j}\hat{H}_{ij})}]^{K+1}$, where
$\beta = (K+1) \tau$, and $\tau$ is the infinitesimal imaginary time
slice. Define a local evolution operator $\hat{U}_{ij} = \exp(-\tau
\hat{H}_{ij})$. Then, the density operator can be represented as
$\rho \simeq [\prod_{i,j} \hat{U}_{ij}]^{K+1} = \prod_{q=1}^{K+1}
\prod_{i,j} \hat{U}^{q}_{ij}$, where $q$ is the Trotter index. By
making a singular value decomposition (SVD) on $U^{ij}_{i'j'} =
\langle ij|\hat{U}_{ij}|i'j'\rangle $ where $|ij\rangle$ stands for
the direct product basis of spins at site $i$ and $j$, we have
$U^{ij}_{i'j'} = \sum_{g} G^{L}_{ii',g} \lambda^{0}_{g}
G^{R}_{jj',g}$, where $\lambda^{0}$ is the singular value vector,
and $G^{L}$ and $G^{R}$ are two local evolution tensors, each of
which has two physical bonds ($i,i'$ and $j,j'$, respectively) and
one geometrical bond ($g$). For a honeycomb lattice, this step is
depicted in Figs. \ref{fig-TensorNet} (a) and (b).

Next, by contracting the shared bonds among $G^{L}$ and $G^{R}$ [Fig.
\ref{fig-TensorNet} (b)], we get
\begin{eqnarray}
 T^{L}_{il,g_1g_2g_3}=\sum_{jk} G^{L}_{ij,g_1} G^{L}_{jk,g_2} G^{L}_{kl,g_3}, \nonumber \\
 T^{R}_{il,g_1g_2g_3}=\sum_{jk} G^{R}_{ij,g_1} G^{R}_{jk,g_2} G^{R}_{kl,g_3},
\label{eq-InitialT}
\end{eqnarray}
where $g_1$, $g_2$ and $g_3$ are three inequivalent bonds on a honeycomb
lattice [Fig. \ref{fig-TensorNet} (c)]. The density operator $\rho$
at an inverse temperature $\tau$ has the form of a TN as
\begin{eqnarray}
 \rho_{\cdots ii'jj' \cdots}=Tr_{G} ( \cdots \lambda^{II}_{g_2} \lambda^{III}_{g_3} T^{L}_{ii',g_1g_2g_3}
 \lambda^{I}_{g_1} T^{R}_{jj',g_1g_2'g_3'} \lambda^{II}_{g_2'} \lambda^{III}_{g_3'} \cdots),
\label{eq-Initialrho}
\end{eqnarray}
in which $Tr_{G}$ is the trace over all contracted geometrical
bonds, and $\lambda^{I}$, $\lambda^{II}$, $\lambda^{III}$ are three
inequivalent singular value vectors with the initial value
$\lambda^{0}$. This gives a TPDO, which is an extension of the
matrix product density operator \cite{MPDO} and the tensor product
states. In fact, the TPDO is an open TN comprised of the infinite
product of two inequivalent tensors $T^{L}$ and $T^{R}$ for two
sublattices of the honeycomb lattice as well as $\lambda^{I}$,
$\lambda^{II}$ and $\lambda^{III}$ for three inequivalent bonds
[Fig. \ref{fig-TensorNet} (c)]. The TPDO at finite temperature
consists of two parts: the initial TPDO and the 3D brick-wall
structure formed by the product of evolution tensors. We can
contract linearly the evolution tensors in pairs into the TPDO along
the imaginary time direction. For example for bond $g_1$, we have
\begin{eqnarray}
 \widetilde{T}^{L}_{ik,(g_1g_1')g_2g_3}=\sum_{j} G^{L}_{ij,g_1'} T^{L}_{jk,g_1g_2g_3},\nonumber \\
 \widetilde{T}^{R}_{ik,(g_1g_1')g_2g_3}=\sum_{j} G^{R}_{ij,g_1'} T^{R}_{jk,g_1g_2g_3},
\label{eq-Evolve}
\end{eqnarray}
and meanwhile, we get
$\widetilde{\lambda}^{I}_{g_1g_1'}=\lambda^{0}_{g_1'}
\lambda^{I}_{g_1}$. The contractions for bonds $g_2$ and $g_3$ are
similar. After certain times of contraction, we obtain the TPDO at
the corresponding inverse temperature. Then by tracing all bonds, we
can get the partition function $Z$ at finite temperature. During the
contraction, as the dimension of the geometrical bonds is
unavoidably enlarged, an optimal approximation is needed to bound
the bond dimension. In the existing algorithms for truncating the
bond, a matrix SVD on the matricization of the tensor is used and the states with $D_c$ (dimension cut-off) largest singular values are preserved.
We extend the Tucker decomposition to the TN's and suggest the NTD to transform the TN into a super-orthogonal form, with which the optimal approximation can be obtained with the robust network singular value spectrum (NSS).

\section{Super-orthogonal form of tensor networks and network Tucker
decomposition}

In the areas of data compression, image processing, etc., Tucker decomposition has been accepted as a convincing higher-order generalization of matrix singular value decomposition, and its approximation scheme for a single tensor has wide and successful applications \cite{TD_App}. It can be written as the product of the form
\begin{eqnarray}
T_{i_1i_2\cdots i_n}=\sum_{j_1j_2\cdots j_n}S_{j_1j_2\cdots j_n} U^{(1)}_{i_1j_1} U^{(2)}_{i_2j_2} \cdots U^{(n)}_{i_nj_n}.
\label{eq-TD}
\end{eqnarray}
The tensor $S$ is called the core tensor and $U^{(k)}_{i_kj_k}$ is
the unitary matrix. This decomposition is considered as a
higher-order generalization of the matrix SVD when the core tensor
$S$ satisfies the following two conditions:

(a) All-orthogonal: $\sum_{i_1i_2\cdots i_{\alpha-1} i_{\alpha+1}
\cdots i_n} S_{i_1 i_2 \cdots i_{\alpha} \cdots i_n} S_{i_1i_2\cdots
i'_{\alpha} \cdots i_n}=0$ if $i_{\alpha} \neq i'_{\alpha}$ for any
$\alpha$;

(b) Ordering: $\parallel S_{i_{\alpha}=1} \parallel \geq \parallel
S_{i_{\alpha}=2} \parallel \geq \cdots \geq \parallel
S_{i_{\alpha}=I_n} \parallel$, where $I_n$ is the dimension of the
index $i_n$, and the norm of the sub-tensor $\parallel
S_{i_{\alpha}=k} \parallel = \sum_{i_1i_2 \cdots i_{\alpha-1}
i_{\alpha+1} \cdots i_n} S_{i_1i_2 \cdots i_{\alpha-1} k
i_{\alpha+1}} S_{i_1i_2 \cdots i_{\alpha-1} k i_{\alpha+1}}^*$.

All-orthogonality requires that each slice (that means fixing one
index when setting others as one composite index free) of the core
tensor $S$ is mutually orthogonal with respect to the scalar product
of matrices. The ordering condition guarantees that the norm of each
sub-tensors of $S$ does not increase as the corresponding index
increases, which is similar to the order of matrix singular values.
Actually, $\parallel S_{i_{\alpha}} \parallel$ is the singular
values of the matrix $M_{ji_{\alpha}}=T_{i_1i_2\cdots i_{\alpha}
\cdots i_n}$ where the composite index $j=(i_1i_2\cdots i_{\alpha-1}
i_{\alpha+1} \cdots i_n)$.

In the Tucker decomposition, the information of
the weight is stored in the core tensor, and more specifically, it
is the norm of each sub-tensor of $S$. The optimal lower-dimensional
approximation of a single tensor can thus be obtained by keeping the
space corresponding to the sub-tensors with larger norms. Several
algorithms of the Tucker decomposition
have been proposed, such as the higher-order orthogonal iteration in
which the interplay among all bonds of the tensor is considered for the
optimal approximation.

In the following, we extend the definition and the approximation scheme of the Tucker decomposition for a single tensor to a TN. First we define the network reduced
matrix (NRM) $\mathcal{M}$ of bond $g_i$ for a (real) tensor $T$ as
\begin{eqnarray}
\begin{aligned}
\mathcal{M}_{g_{i}g'_{i}}&=\sum_{p} \sum_{g_{1}g_{2}\cdots g_{n}} T_{p,g_{1}g_{2}\cdots g_{i}\cdots g_{n} }
T_{p,g_{1}g_{2}\cdots g'_{i}\cdots g_{n}} (\lambda_{g_{1}}\lambda_{g_{2}}\\ & \cdots \lambda_{g_{i-1}}
\lambda_{g_{i+1}}\cdots \lambda_{g_{n}})^2 \lambda_{g_{i}} \lambda_{g'_{i}},
\label{eq-NRM}
\end{aligned}
\end{eqnarray}
where $p=\{p_{1},p_{2},\cdots ,p_{m}\}$ denotes the composite bond
of all physical indices because one can always rearrange all
physical indices into a composite index, $g_{i}$ denotes a
geometrical bond, and $T_{p,g_{1}g_{2}\cdots g_{i}\cdots g_{n} }$
represents an element of the tensor $T$. The super-orthogonal form
of an open TN is defined by two conditions:

(a) Ordering: all $\lambda$'s on geometrical bonds are positive-defined, normalized
and the elements of each $\lambda$ are in descending order. We coin
the vectors $\lambda$'s as the network singular value spectrum, that
is a generalization of the matrix singular value spectrum.

(b) Orthogonality: for any tensor $T$ in the TN and any
geometrical index $g_{i}$ of $T$, the NRM $\mathcal{M}$ is diagonal
and equals to the square of the corresponding $\lambda$, say
$\mathcal{M}_{g_{i}g'_{i}} = \lambda_{g_{i}}^2
\delta_{g_{i}g'_{i}}$.

The super-orthogonal conditions, which are nonlocal, require that
the matrix $A_{\{pg_{1}g_{2}\cdots g_{i-1}g_{i+1}\cdots
g_{n}\},g_{i}}=T_{p,g_{1}g_{2}\cdots
g_{n}}\lambda_{g_{1}}\lambda_{g_{2}}\cdots
\lambda_{g_{i-1}}\lambda_{g_{i+1}}\cdots \lambda_{g_{n}}$ (that is
analog to the singular vectors of matrix SVD) is column orthogonal
for any $i$. These conditions are global constraints for the TN, as
every tensor should satisfy them simultaneously. From the NSS which
contains the information of the weight distribution instead of the core tensor in the Tucker decomposition, the optimal low-dimensional approximation of the bond space can be obtained.
For 1D systems, the super-orthogonal conditions require that the
matrix product states (MPS)'s satisfy simultaneously the left and
right canonical conditions defined in Ref. [\onlinecite{Canonical}],
which leads to the canonical form of MPS.

\begin{figure}[tbp]
\includegraphics[angle=0,width=1\linewidth] {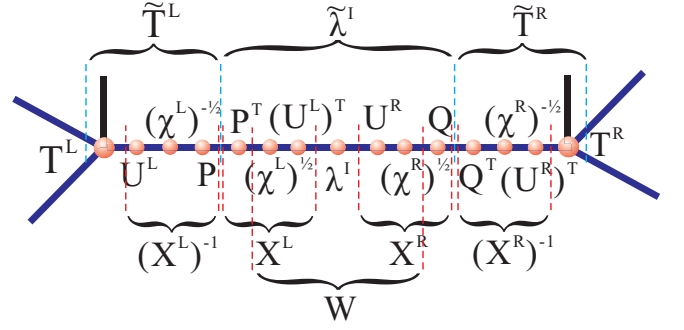}
\caption{(Color online) The identical transformation within the NTD
for the bond $g_1$ of an open honeycomb TN.}
\label{fig-Transformation}
\end{figure}

Within the suggested NTD, the super-orthogonal form is gained by
iteratively transforming the TN with identical transformations on
each geometrical bond until the pre-established convergence to the
super-orthogonal form is reached. For instance, for bond $g_1$ (Fig.
\ref{fig-Transformation}) the transformation matrices $X^L$ and
$X^R$ for $T^L$ and $T^R$ (the transformation matrices $Y^L$ and
$Y^R$ for bond $g_2$, $Z^L$ and $Z^R$ for bond $g_3$ are similar)
are defined by

\begin{eqnarray}
X^L_{ab}=\sum_{c} P_{ca} (\chi^L)^{-1/2}_{c} U^L_{bc},\nonumber \\
X^R_{ab}=\sum_{c} Q_{ca} (\chi^R)^{-1/2}_{c} U^R_{bc},
\label{eq-TransfM}
\end{eqnarray}
where $U^{L(R)}_{bc}$ contains eigenvectors and $\chi^{L(R)}_c$
contains the eigenvalues of the matrix
$\overline{M}^{L(R)}_{bc}=(\lambda^{I}_{b} \lambda^{I}_{c})^{-1}
\mathcal{M}^{L(R)}_{bc}$. $P$ ($Q$) contains the left (right)
singular vectors of the intermediate matrix $W$ defined by

\begin{eqnarray}
W_{ac}=\sum_{b} (\chi^{L})^{1/2}_{a} U^{L}_{ba}
\lambda^I_{b} U^{R}_{bc} (\chi^R)^{1/2}_{c} \overset{svd}= \sum_{g} P_{ag} \widetilde{\lambda}^I_{g} Q_{bg}.
\label{eq-IntermediateM}
\end{eqnarray}
By inserting the identity $I=X^L(X^L)^{-1}=X^R(X^R)^{-1}$ into the
left (right) side of $\lambda^I$ and transforming $T^L$, $T^R$ and
$\lambda^I$, we have

\begin{eqnarray}
\widetilde{T}^{L}_{p,gbc}=\sum_{a} T^{L}_{p,abc} (X^L)^{-1}_{ag}; \quad
\widetilde{T}^{R}_{p,gbc}=\sum_{a} T^{R}_{p,abc} (X^R)^{-1}_{ag};\\
\widetilde{\lambda}^I_{g} \delta_{gg'}=\sum_{a} X^{L}_{ga} \lambda^I_{a} X^{R}_{ag'}= \sum_{af} P_{ag} W_{af} Q_{fg'}.
\label{eq-NewT}
\end{eqnarray}
With new $\widetilde{T}^{L}$, $\widetilde{T}^{R}$ and
$\widetilde{\lambda}^I$ as well as $\lambda^{II}$ and
$\lambda^{III}$, the NRM of the bond $g_1$ equals to
$(\widetilde{\lambda}^{I})^{2}$ as the matrices
$\overline{A}^{L(R)}_{p,{abc}}=\sum_{a'} T^{L(R)}_{p,a'bc}
\lambda^{II}_{b} \lambda^{III}_{c} U^{L(R)}_{a'a}
(\chi^{L(R)})^{-1/2}_{a}$ and $P(Q)$ are both column orthogonal and
normalized.

After doing similar transformations on the other two bonds,
$\overline{A}$ bears the form of

\begin{eqnarray}
\overline{A}^{L(R)}_{p,{abc}}=\sum_{a'} \widetilde{T}^{L(R)}_{p,a'bc}
\widetilde{\lambda}^{II}_{b} \widetilde{\lambda}^{III}_{c} U^{L(R)}_{a'a} (\chi^{L(R)})^{-1/2}_{a} \nonumber \\
=\sum_{a'b'c'} T^{L(R)}_{p,a'b'c'} (\lambda^{II}_{b'} Y^{R(L)}_{b'b})
(\lambda^{III}_{c} Z^{R(L)}_{c'c}) U^{L(R)}_{a'a} (\chi^{L(R)})^{-1/2}_{a}.
\label{eq-Tbar}
\end{eqnarray}
It can be seen from Eqs. (\ref{eq-TransfM}) and (\ref{eq-Tbar}) that
the super-orthogonal conditions are satisfied when the eigenvalues
$\chi^{L(R)}$ in Eq. (\ref{eq-TransfM}) are uniformly distributed
(i.e., all eigenvalues are equal to $1$) for all three bonds. The
deviation of $\chi^{L(R)}$ from uniform distribution can be measured by
$\zeta=(|\chi^{L}-\mathcal{V}|+|\chi^{R}-\mathcal{V}|)/(2\mathcal{L})$, in which $|\bullet|$
means the norm of a vector, $\mathcal{V}$ is a vector with all its
elements equal to 1 and $\mathcal{L}$ is the length of the vector
$\chi$. In addition, we define a factor that
measures the convergence of $\lambda$'s at the $t$th iteration by
$\mu(t)=\sum_{S=I,II,III} (|\lambda^{S}(t-3) - \lambda^{S}(t)| /
|\lambda^{S}(t)|)/3$, and a factor $\sigma = (\sigma^L +
\sigma^R)/2$, where $\sigma^{L(R)}=\sum_{ab}
|\mathcal{M}^{L(R)}_{ab}-(\lambda^S_a)^2 \delta_{ab}|$ measures TN's
deviation from the super-orthogonal form according directly to the
super-orthogonal conditions.

\begin{figure}[tbp]
\includegraphics[angle=0,width=1\linewidth]{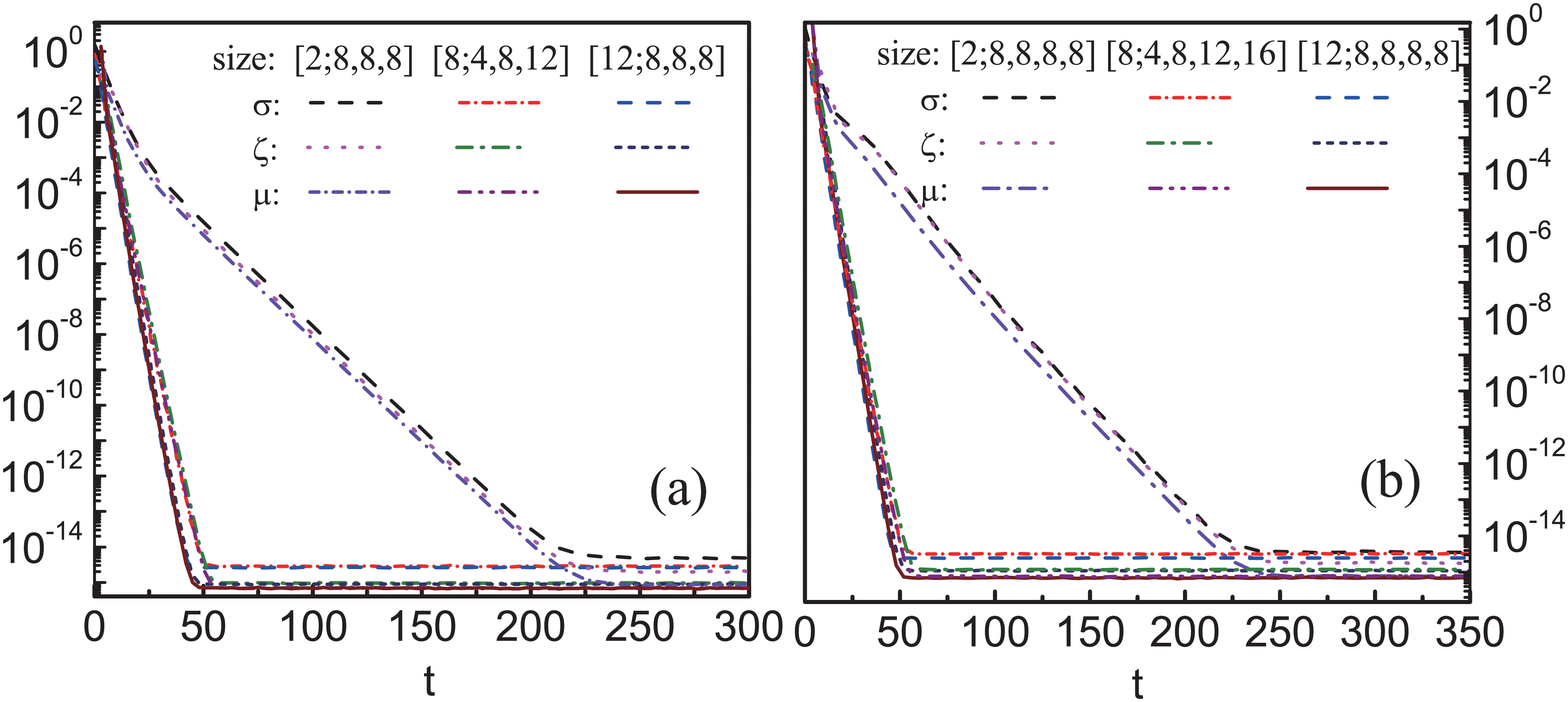}
\caption{(Color online) The convergence of $\mu$, $\sigma$ and
$\zeta$ of open (a) honeycomb and (b) square TN's with the increase
of the iteration step $t$. $[D_p;D_{g_1},D_{g_2},\cdots ,D_{g_n}]$
is the size of the tensors that form the TN, where
$D_{g_i}(i=1,2,\cdots,n)$ is the dimension of the geometrical bonds
and $D_p$ is the dimension of the physical bond. Each factor is
obtained from the average of the results of $100$ randomly
initialized TN's. The error of the eigenvalue decomposition
$\varsigma$ is about $10^{-15}$.} \label{fig-Convergence}
\end{figure}

To testify the robustness of the super-diagonal form and the
efficiency of NTD, we randomly initialize the inequivalent tensors
and vectors (according to Gaussian distribution $N(0,1)$) that form
the infinite open honeycomb and square TN, and calculate the factors
$\mu$, $\sigma$ and $\zeta$ with different iteration steps. The
value of each factor is the average of the results of $100$ randomly
initialized TN's. Fig. \ref{fig-Convergence} shows that the NTD can
transform a randomly initialized TN into a super-orthogonal form
very efficiently. It is found that $\zeta$ and $\sigma$ decay
exponentially to about $10^{-14} \thicksim 10^{-15}$ (where the
error of the eigenvalue decomposition itself $\varsigma$ is about
$10^{-15}$) within $300$ steps. Meanwhile, $\mu$
converges to $10^{-14} \thicksim 10^{-15}$, which justifies the good
convergence of the three inequivalent $\lambda$'s. We may see that
for a certain TN, the three factors share one same
super-orthogonalization ratio $\xi$ and in general, $\xi$ becomes
larger when we increase the space of the physical bond and fix the
space of the geometrical bonds.

The computational cost to transform a TN into the super-orthogonal
form with NTD is mainly from the eigenvalue (singular value)
decompositions of $D_g\times D_g $ matrices, which is about
$O(tD_g^{3})$ with $t$ the transformation steps and $D_{g}$ the
dimension of the geometrical bond. In the ODTNS scheme, the TPDO
converges to the super-orthogonal form with $\sigma < 10^{-8}$ only
with $t \thicksim 10$ steps, because for a small $\tau$, the
evolution is nearly identical.

\section{The free energy}

After each time that the evolution tensors
are contracted into the TPDO, we super-orthogonalize the TPDO and
obtain the optimal approximation of the enlarged geometrical bonds
as well as the $\lambda^S$'s ($S=I,II,III$). By collecting the
normalization factor $r^S_q = \sqrt{ \sum_{g} \lambda^S_g }$ with
$q$ the Trotter step, the free energy per site can be obtained with
$r^S_q$ and $\overline{r}$ \cite{LastLayer} that is the contraction
of the TPDO by
\begin{eqnarray}
 f(\beta)=\frac{1}{2\beta}(\sum^{K}_{q=1}\sum_{S=I,II,III}\ln{r^{S}_{q}}+
 2\ln{\overline{r}}).
\label{eq-FreeE}
\end{eqnarray}
The thermodynamic quantities of the 2D quantum lattice systems can
be obtained from the free energy $f(\beta)$.

What is more, the ground state properties can also be obtained with
the ODTNS scheme. When one takes $K\rightarrow \infty$ and $\tau
\rightarrow 0$, the ground state energy per site $e_0$ has a simple
form of
\begin{eqnarray}
 e_{0}=\lim_{K\rightarrow \infty}\lim_{\tau\rightarrow0}\frac{1}{2\tau} \ln \prod_{S=I, II, III}{{r}^{S}}.
\label{eq-FreeEg}
\end{eqnarray}

\section{Thermodynamics of spin-1/2 antiferromagnet on a honeycomb
lattice}

To judge the efficiency and accuracy of the ODTNS
algorithm, let us consider the spin-1/2 antiferromagnet on an
infinite honeycomb lattice with $\hat{H}_{ij}=\delta(\hat{S}^{x}_{i}
\hat{S}^{x}_{j} + \hat{S}^{y}_{i} \hat{S}^{y}_{j}) +\hat{S}^{z}_{i}
\hat{S}^{z}_{j}$, where $\delta$ measures the anisotropy of spin
interactions. In following calculations, $\mu$ is kept smaller than
$10^{-8}$ \cite{mu}, and the lattice size for QMC calculations is
$64\times64$.

\begin{figure}[tbp]
\includegraphics[angle=0,width=1\linewidth] {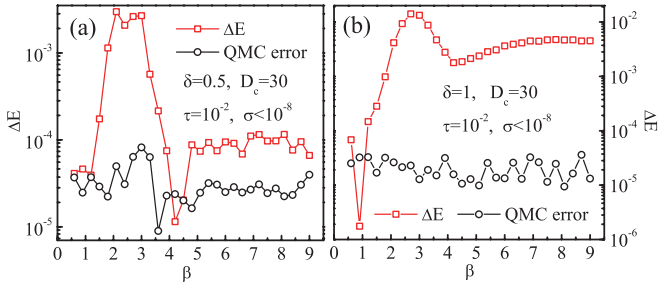}
\caption{(Color online) The energy difference $\Delta E=|E-E_{QMC}|$
between the ODTNS and QMC calculations and the QMC error at
different inverse temperature $\beta$ for the spin-1/2 Heisenberg
antiferromagnet on a honeycomb lattice. (a) shows the result at
$\delta=0.5$ when the system suffers a thermodynamic phase
transition; (b) shows the result at $\delta=1$ when the system is
gapless and a thermodynamic phase transition is forbidden by
Mermin-Wagner theorem. We set $\tau=10^{-2}$, $\sigma<10^{-8}$ and
$D_c=30$.} \label{fig-beta}
\end{figure}

Fig. \ref{fig-beta} shows the energy difference $\Delta
E=|E-E_{QMC}|$ between the results obtained by ODTNS and QMC
calculations at different inverse temperature $\beta$ in the absence
of the magnetic field, where $\tau=10^{-2}$ and $D_c=30$. We find
that, when $\delta=0.5$, there exists a thermodynamic phase
transition, and the energy difference is about $10^{-4} \thicksim
10^{-5}$ at both high and low $\beta$. Near the critical point the
difference is relatively high but is still smaller than $0.003$.
When $\delta=1$, the system is gapless, and the energy difference is
about $10^{-3}$ at both high and low $\beta$, and near the crossover
point, the difference is also relatively high but still remains
around $10^{-2}$. These results show that the precision of ODTNS
scheme is comparable with that of QMC.

\begin{figure}[tbp]
\includegraphics[angle=0,width=1\linewidth] {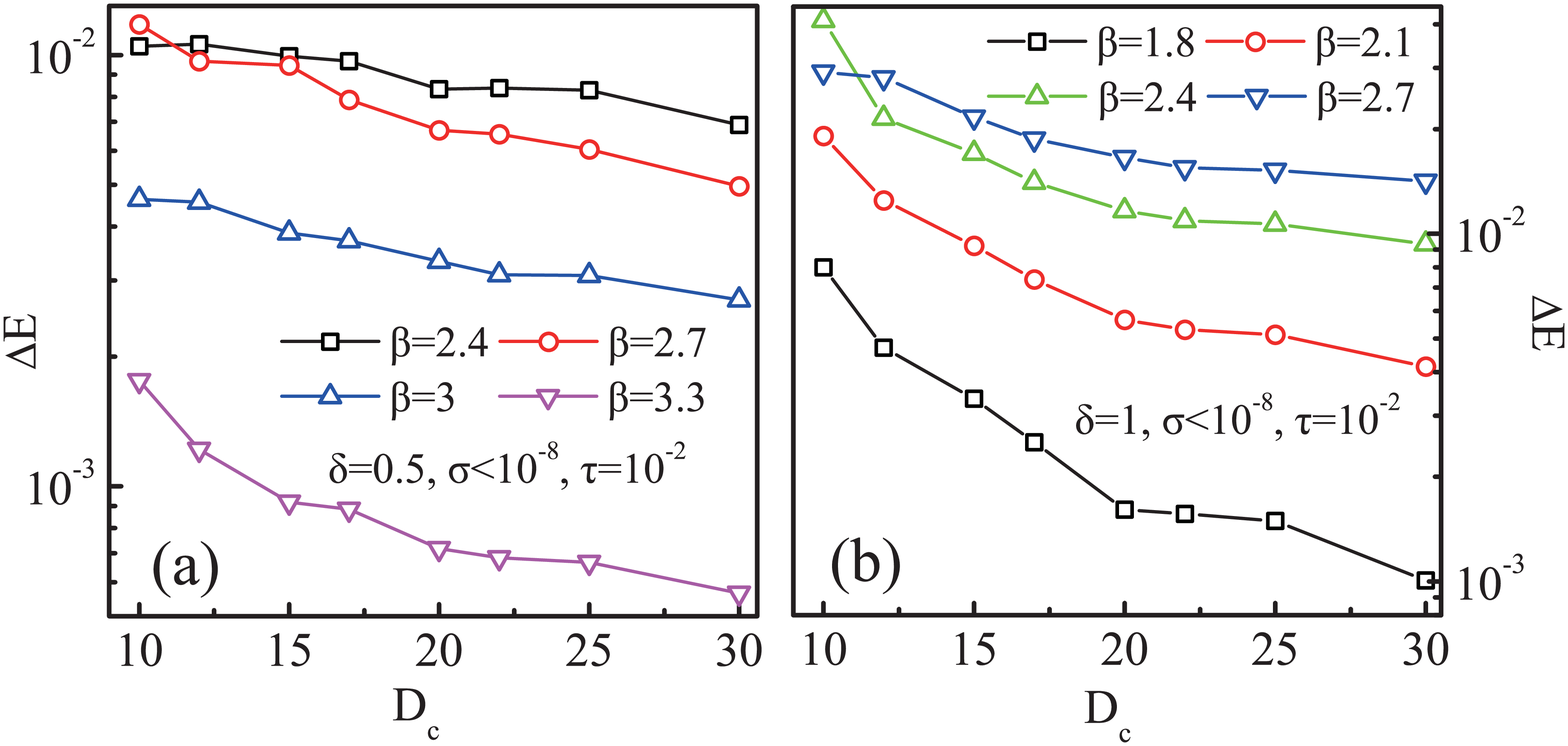}
\caption{(Color online) The energy difference $\Delta E=|E-E_{QMC}|$
between ODTNS and QMC with the dimension cut-off $D_c$ near the
critical (crossover) point for (a) $\delta=0.5$ and (b) $\delta=1$.
It can be seen that $\Delta E$ becomes smaller as $D_c$ is
increased. The QMC error is around $10^{-5}$.} \label{fig-Dc}
\end{figure}

We investigated $\Delta E$ versus the dimension cut-off $D_c$ near
the critical (crossover) $\beta$, as shown in Fig. \ref{fig-Dc},
where $\tau=10^{-2}$. It is observed that the energy difference
becomes smaller when $D_c$ is increased. When $\beta$ is away from
the critical (crossover) point, we uncovered that different $D_c$
gives errors within $10^{-4}$. We also checked $\Delta E$ for
different $\tau$, and disclosed the (Trotter) errors are within
$10^{-4}$.

\begin{figure}[tbp]
\includegraphics[angle=0,width=1\linewidth] {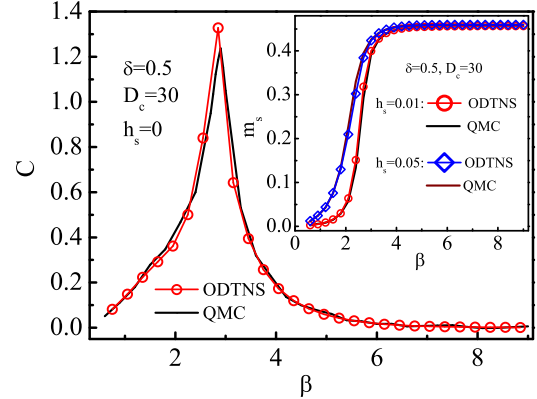}
\caption{(Color online) The inverse temperature $\beta$ dependence
of the specific heat at $\delta=0.5$ and $h_s=0$ for the spin-$1/2$
anisotropic Heisenberg antiferromagnet on a honeycomb lattice. The
QMC result with the error around $10^{-3}$ is included for a
comparison. The inset shows the staggered magnetization at the
magnetic field $h_s=0.01$ and $0.05$, and the QMC error is around
$10^{-5}$.} \label{fig-transition}
\end{figure}

The specific heat as a function of $\beta$ is calculated by
$C=-\beta^{2}dE/d\beta$, as shown in Fig. \ref{fig-transition} for
$\delta=0.5$. A divergent peak at a critical temperature $T_c$ is
observed, which indicates that a phase transition occurs between a
paramagnetic phase and an antiferromagnetic phase at $T_c$. Such a
phase transition is also confirmed with the result of the staggered
magnetization, shown in the inset of Fig. \ref{fig-transition}. The
QMC results are also included for a comparison. One may see that
both results from the ODTNS and QMC calculations agree quite well,
showing again the efficiency and accuracy of the present method. In
addition, the present ODTNS algorithm can be directly applied to the
2D frustrated quantum spin models.

\section{Thermodynamics of a spin-1/2 frustrated bilayer honeycomb Heisenberg model}

\begin{figure}[tbp]
\includegraphics[angle=0,width=1\linewidth]{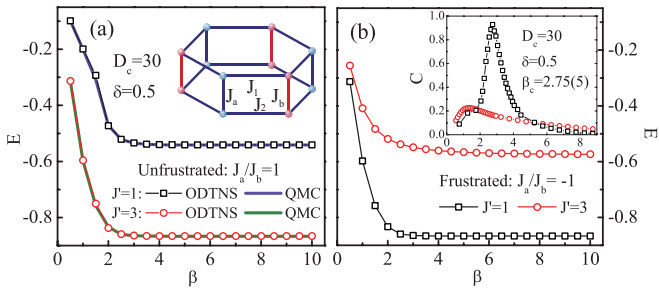}
\caption{(Color online) The inverse temperature $\beta$ dependence
of the energy for the spin-$1/2$ Heisenberg model on a bilayer
honeycomb lattice for (a) $J'=J_a=J_b=1$ and $3$, and (b)
$J'=J_a=-J_b=1$ and $3$. The inset of (a) shows the bilayer
structure of the system and the inset of (b) shows the specific heat
$C$ of the 2D frustrated quantum spin system.} \label{fig-Bilayer_E}
\end{figure}

We apply the ODTNS algorithm to explore the spin-$1/2$ Heisenberg model on a bilayer honeycomb
lattice with the Hamiltonian $\hat{H}=\sum_{<ij>} \hat{H}_{ij}$, where
$\hat{H}_{ij}=\hat{H}_{ij}^{(1)} + \hat{H}_{ij}^{(2)} +
(\hat{H}_{i}^{(a)} + \hat{H}_{j}^{(b)})/3$,
$\hat{H}_{ij}^{(1,2)}=J_{1,2} [\delta_{1,2} (\hat{S}^x_i \hat{S}^x_j
+ \hat{S}^y_i \hat{S}^y_j) + \hat{S}^z_i \hat{S}^z_j]$ the
anisotropic Heisenberg antiferromagnet on each layer and
$\hat{H}_{i}^{(a,b)} = J_{a,b} [\delta_{a,b} (\hat{S}^x_i
\hat{S}^x_j + \hat{S}^y_i \hat{S}^y_j) + \hat{S}^z_i \hat{S}^z_j]$
the interlayer coupling [see the inset of Fig. \ref{fig-Bilayer_E}
(a) for the layout of the model], $\delta_{1,2}$ and
$\delta_{a,b}$ measure the corresponding anisotropy of nearest
neighbor spin interactions. We take $J_1=J_2=1$ as energy scale.

When both $J_a$ and $J_b$ are positive, the couplings are all
antiferromagnetic, and the system has no frustration. The
energies obtained by the ODTNS algorithm at $J'=J_a=J_b=1$ and $3$
are shown in Fig. \ref{fig-Bilayer_E} (a), which are in
good agreement with QMC results (where the QMC error is within
$10^{-5}$).

\begin{figure}[tbp]
\includegraphics[angle=0,width=1\linewidth]{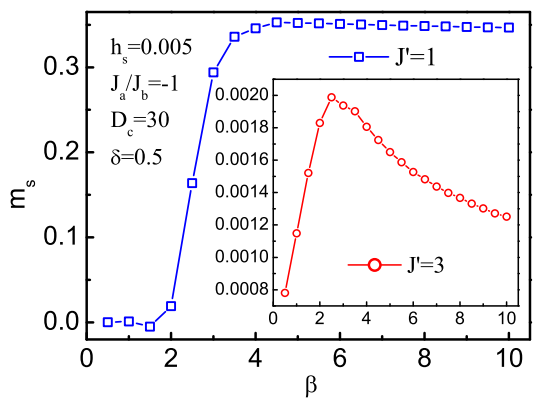}
\caption{(Color online) The sublattice magnetization per site $m_s$
at $J'=J_a=1$ and $3$ (inset) for the spin-$1/2$ frustrated
Heisenberg model on a bilayer honeycomb lattice, where
$J_a/J_b=-1$.} \label{fig-Bilayer_ms}
\end{figure}

When $J_a$ and $J_b$ take different signs, the system
becomes frustrated and the QMC simulations fail because of suffering from the negative
sign problem. Fig. \ref{fig-Bilayer_E} (b) shows the results
of energy and specific heat at $J'=J_a=-J_b=1$ and $3$.
When $J'=J_a=1$, a second-order phase
transition is found at $\beta_c=2.75(5)$; when $J'=3$, no phase
transition is observed from the specific heat. Notice that the
frustration reaches the maximum at $J'=3$ in the Ising limit for
this present model. Fig. \ref{fig-Bilayer_ms} presents the
sublattice magnetization per site $m_{s}$ at $J'=1$ and $3$ where
the frustration exists. At $J'=1$, the couplings within each layer
are dominant. The system is in the antiferromagnetic phase at
low temperature and there exists a thermal phase transition from the
antiferromagnetic to paramagnetic phase. At $J'=3$ when the strong
frustration is present, $m_s$ is around $10^{-3}$, indicating the
absence of magnetic long range order at all temperature. These
calculations show that the ODTNS algorithm is capable of studying the 2D
frustrated quantum spin systems.

\section{Summary}

In summary, a novel algorithm based on the ODTNS scheme for the 2D
quantum spin lattice models is proposed. By mapping the 2D quantum
model into a 3D TN, we suggest the NTD to obtain the optimal
approximation of the bond space by transforming the TPDO into the
super-orthogonal form, that leads to an efficient and accurate
calculation of the free energy as well as other observables in the
2D quantum systems. We testify the efficiency and accuracy of the
present algorithm by studying the thermodynamics of a spin-$1/2$
Heisenberg antiferromagnet on a honeycomb lattice, and compare the
results with those of the QMC. It is shown that the precision of the ODTNS algorithm is comparable
with that of the QMC as both results agree very well. In addition, we find that the present
algorithm can also be applied to explore the 2D frustrated quantum spin
models without suffering from a negative-sign problem. It is
expected that the present ODTNS scheme could also be extended to 2D
correlated electron systems.

\acknowledgements

The authors are indebted to J. Chen, X. Yan, F. Yei, Y. Zhao and Q. R. Zheng
for stimulating discussions. This work is supported in part by the
NSFC (Grant Nos. 90922033 and 10934008), the MOST of China (Grant
No. 2012CB932901) and the CAS.


\begin{thebibliography}{99}
\bibitem{LastLayer} We contract the TPDO by first tracing the physical bonds to obtain an infinite 2D closed TN and then contracting this TN with the ODTNS algorithm.
\bibitem{mu} Note that without the iteration in the NTD,
$\sigma$ is on the same order of $\tau$ for the imaginary time
evolution, which could lead to the results unstable near the
critical (crossover) point. If one reduces $\tau$ to decrease
$\sigma$ without making iterations in the NTD, the cost would be
very high because the extremely large evolution steps are needed to
reach a certain temperature.
\end{thebibliography}
\end{document}